# Arbitrary Shaped Acoustic Concentrators Enabled by Nihility Media

**Mohammad Hosein Fakheri, Hooman Barati Sedeh[1] and Ali Abdolali[1,*]**

[1] Applied Electromagnetic Laboratory, School of Electrical Engineering, Iran University of Science and Technology, Tehran, 1684613114, Iran
[*]corresponding: Abdolali@.iust.ac.ir
[+]These authors contributed equally to this work

## Abstract

Based on the transformation acoustic (TA) methodology, an innovative approach for designing arbitrary shape concentrators is proposed. Unlike previous works, which utilized inhomogeneous and anisotropic materials to localize the incident acoustic waves in an arbitrary domain, the same functionality will be attained by introducing only one homogeneous anisotropic medium, which is called Acoustic nihility media (ANM). A great advantage of this method is that the attained materials are not dependent on the shape of the concentrator. That is regardless of the device geometry, a constant ANM will be used for each new shape and the output results do not alter. This will circumvent the conventional transformation acoustics' sophisticated and tedious calculations and could be easily implemented in real-life scenarios.

## Introduction

Recently, manipulation of acoustic waves has gained lots of attention, due to its several applications in sonar detection [1], ultrasonic imaging [2] ultrasonic examination [3] and geological research [4]. Among these applications, the phenomena of the near-field concentration of sound waves play an important role in energy harvesting and increasing the acoustic intensity in a specific volume of space. For instance, in the fields of power ultrasonic and underwater sound, acoustic concentrators are widely used in order to magnify vibrational displacement amplitude, transform vibrational direction and to focus energy [5]. Furthermore, in the underwater imaging sonar systems, an acoustic lens, which functions as an acoustic concentrator, is used to make high sound pressure in a small region of interest in order to increase the image resolution of the target [6]. Although the results of these conventional methods are fascinating, the complexity of the design procedure and the intricacy of the structures, limit their practical implementation in real-life scenarios. Therefore, the requirement of a method for designing arbitrary shape concentrators, which are puissant to be designed and fabricated easily, was highly felt. Fortunately, the recently proposed concept of transformation acoustic (TA) proved its usefulness for this aim.

TA methodology, which was proposed by Cummer *et al.* [7] is based on the fact that the manipulation of acoustic waves is caused by the materials, which are mainly anisotropic and inhomogeneous. It was shown that these materials are achieved via transforming an original domain (named as virtual space) into the transformed domain (expressed by physical space) with an appropriate mapping function [7]. Since the introduction of this innovative method, many novel acoustic devices that were deemed impossible to be achieved such as acoustic rotators [8], illusion devices [9], acoustic beam collimating lenses [10] and acoustic cloaks[11] have been realized and experimentally demonstrated with acoustic metamaterials. In addition to the mentioned applications, several TA-based devices have been reported for focusing the acoustic waves in a specific volume of space [12-15]. However, to the authors' best knowledge, most of the previous works were limited to the cylindrical-cross section concentrators which limit their performances for being used in more general cases where there is a prerequisite to focus acoustic intensity in an arbitrary region of interest. Although Yang *et al.* [15] performed some theoretical investigations for this aim, the propounded approach cannot be utilized in realistic situations since its obtained materials are geometry dependent. That is by changing the geometry of the transformed medium, one must recalculate its necessitating materials, which is a tedious and sophisticated procedure especially when the geometry of the preferable media is irregular. More recently, a feasible approach based on quasi-conformal mapping (QCM) has been exploited to achieve acoustic devices with only inhomogeneous materials [16]. Albeit, not only the intricate mathematical calculations during the design process is the main drawback of this approach, but also the functionality of the attained devices are extremely restricted to some specific shapes. Therefore, in the light of the above-mentioned reasons, the question is raised that whether is there a much simpler way of designing acoustic concentrators that their necessitating materials are independent of the device geometry? In other words, is it possible to use a simple material for any arbitrary irregular shape concentrators and achieve perfect focusing functionalities?

In this paper, a new method to attain material parameter expressions for any desired shape concentrators is proposed that is more feasible to be implemented in the real-life scenario. Due to the presented technique, the acquired materials became acoustic nihility media (also known as acoustic null materials or(ANM)) which is recently exploited by Sun. et al. to design acoustics multi emission lenses[17]. In contrary to the conventional TA based materials, which must be recalculated when the shape of the structure is changed, ANM are independent of the structure geometry. This, in turn, will give the designer a more degree of freedom to arbitrarily manipulate the desired wave. We believe that not only this method could find applications in scenarios where high field pressure in an arbitrary region is highly demanded such as in power ultrasonics applications, underwater sound manipulation and sonar systems but also it is more feasible to be fabricated in comparison with its conventional counterparts.

## Design Principals

In order to achieve the perfect functionality of a concentrator, the schematic diagram of Fig.1(a) is used as the space transformation. As it can be seen, three cylinders with arbitrary cross sections enclosed by contours $R_1(\phi)$, $R_2(\phi)$ and $R_3(\phi)$ divide the space into three different regions. To concentrate the incident acoustic wave in a predefined region, one must collect all of the energy that originally locates in $r < R_2(\phi)$ into the region of $r' < R_1(\phi')$. To this aim, a two-steps mapping function of Fig.1 (b) is used as the coordinate transformation.

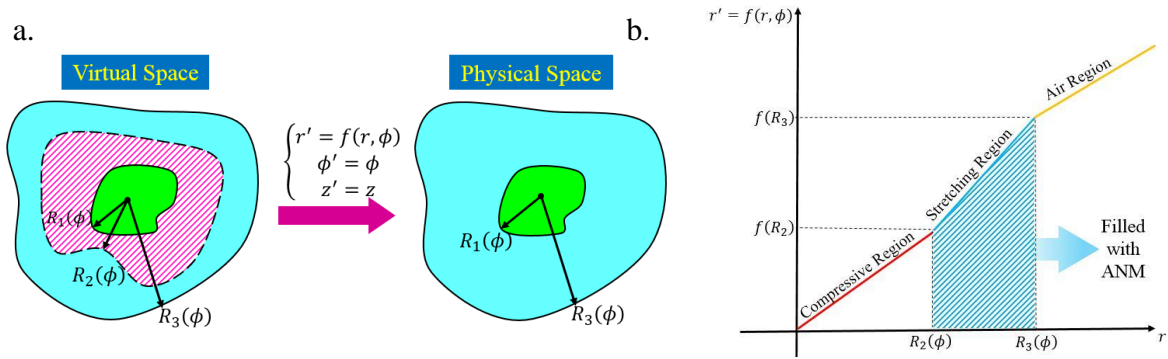

**Fig.1. (a)** The schematic of coordinate transformation for achieving arbitrary shape concentrators. **(b)** Utilized two-steps mapping function

To localize the energy, the region of $r \in [0, R_2(\phi)]$ must be mapped to $r' \in [0, R_1(\phi')]$. To this aim, while the region of $r \in [0, R_2(\phi)]$ is compressed into $r' \in [0, R_1(\phi')]$ ( red line of Fig.1 (b)), region between $r \in [R_2(\phi), R_3(\phi)]$ must be stretched into region $r' \in [R_1(\phi'), R_3(\phi')]$ $ as it is indicated with blue line in the same figure. Since these two steps occur simultaneously, all the energy, which previously located in $r < R_2(\phi)$ is now localized in the region of $r' < R_1(\phi')$ and as a result, the acoustic intensity will be increased in the mentioned domain. In addition, since the proposed transformation does not alter the azimuthal direction ( i.e. $\phi' = \phi$), the final transformation function could be expressed as:

$$\begin{cases} f_c(r,\phi) = \dfrac{R_1(\phi)}{R_2(\phi)} r & r' \in [0, R_1(\phi)) \\ \\ f_s(r,\phi) = \left( \dfrac{R_3(\phi) - R_1(\phi)}{R_3(\phi) - R_2(\phi)} \right) r + \left( \dfrac{R_1(\phi) - R_2(\phi)}{R_3(\phi) - R_2(\phi)} \right) R_3(\phi) & r' \in [R_1(\phi), R_3(\phi)] \end{cases} \quad (1)$$

where the subscript of $c, s$ represents each compressive and stretching regions discussed above. From TA methodology point of view, under any coordinate transformation from original space $(x, y, z)$ to a new space $(x', y', z')$, the mass density and bulk modulus of the physical space are expressed as [7]:

$$\rho' = \rho_0 \det(\Lambda) (\Lambda^{-1})^T (\Lambda^{-1}), \quad \kappa' = \kappa_0 \det(\Lambda) \quad (2)$$

where $\rho_0$ and $\kappa_0$ are the mass density and bulk modulus of the virtual domain which is assumed to be filled with air and $\Lambda$ is the Jacobian matrix denoted by $\Lambda = \partial(x', y', z') / \partial(x, y, z)$. By substituting Eq. (1) into Eq. (2), the necessitating materials will be obtained as:

$$\rho' = \rho_0 \begin{pmatrix} \rho_{11} & \rho_{12} & 0 \\ \rho_{21} & \rho_{22} & 0 \\ 0 & 0 & \rho_{33} \end{pmatrix}, \kappa' = \kappa_0 \frac{f_i(r,\phi) \times \dfrac{\partial f_i(r,\phi)}{\partial r}}{r} \quad (3\text{-a})$$

$$\begin{aligned} \rho_{11} &= \frac{f_i(r,\phi)}{r \times \partial f_i(r,\phi)/\partial r} \\ \rho_{12} &= \rho_{21} = -\frac{\partial f_i(r,\phi)/\partial \phi}{r \times \partial f_i(r,\phi)/\partial r} \\ \rho_{22} &= -\frac{\partial f_i(r,\phi)/\partial \phi}{r \times \partial f_i(r,\phi)/\partial r} \left\{ \left( \frac{\partial f_i(r,\phi)/\partial \phi}{r \times f_i(r,\phi) \times \partial f_i(r,\phi)/\partial r} \right)^2 + \frac{r \times \partial f_i(r,\phi)/\partial r}{f_i(r,\phi)} \right\} \\ \rho_{33} &= \frac{f_i(r,\phi) \times \partial f_i(r,\phi)/\partial r}{r} \end{aligned} \quad (3\text{-b})$$

As it can be seen from Eq.(3), the obtained materials are inhomogeneous and anisotropic with off-diagonal components in both regions, which cause serious difficulties for their fabrication. However, in the practical scenarios only the geometry of the region where the acoustic intensity should be increased in is important; therefore, without the loss of generality, one can assume that these contours are conformal. In other words, $R_1(\phi) = \tau_1 R(\phi)$, $R_2(\phi) = \tau_2 R(\phi)$ and $R_3(\phi) = \tau_3 R(\phi)$ where $\tau_i$ are constant coefficients which satisfied the condition of $\tau_1 < \tau_2 < \tau_3$ and $R(\phi)$ is an arbitrary shape contour as shown in Fig.2(a). Moreover, $R(\phi)$ is an arbitrary continuous function with a period of $2\pi$ that could be expressed via Fourier series in the form of:

$$R(\phi) = a_0 + \sum_{n=1}^{\infty} a_n \cos(n\phi) + b_n \sin(n\phi) \tag{4}$$

where $a_n$ and $b_n$ are constant coefficients that specify the contour shape.

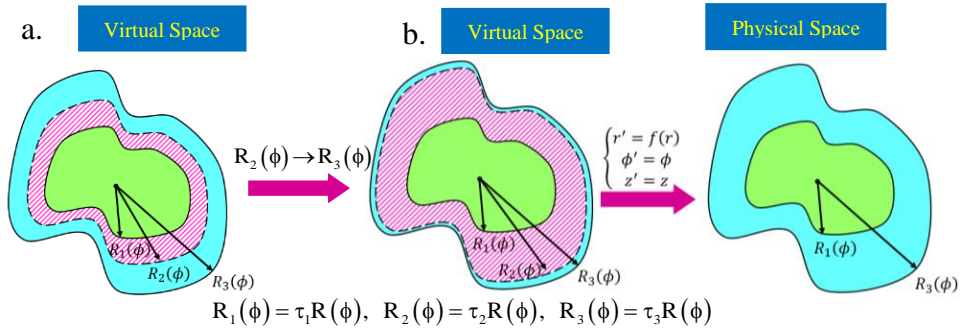

**Fig.2. (a)** The schematic of coordinate transformation for achieving arbitrary shape concentrators with conformal boundaries. **(b)** Coordinate transformation when $R_2(\phi) \rightarrow R_3(\phi)$

This simplifying assumption can leads to more simplified materials with the obviated off-diagonal components in the comprehensive section, since $f_c(r,\phi)$ would be independent of $\phi$ in this region (i.e. $f_c(r) = (\tau_1/\tau_2)r$ ).Nevertheless, in the stretching region the off-diagonal components will remain due to the existence of $R(\phi)$ in the second term of $f_s(r,\phi)$. However, as $R_2(\phi) = \tau_2 R(\phi)$ is a fictitious region, $\tau_2$ can achieve any arbitrary value. This will give us a degree of freedom to arbitrarily select the value of $\tau_2$ in a manner that it will eradicate the off-diagonal components of $\rho_{12}$ (also $\rho_{21}$). To this aim, we will set $\tau_2 \rightarrow \tau_3$ (Fig.2(b)), which in turn make $\partial f_s(r,\phi)/\partial r \rightarrow \infty$. According to Eq.(3), this will give rise to abolishing the off-diagonal components in the stretching region (i.e. $\rho_{12} = \rho_{21} \rightarrow 0$). Therefore, after some simple calculations, the inhomogeneous and anisotropic materials of Eq.(3) will be changed to more simplified ones as:

$$\rho'_s = \begin{pmatrix} 0 & 0 & 0 \\ 0 & \infty & 0 \\ 0 & 0 & \infty \end{pmatrix}, \rho'_c = \begin{pmatrix} 1 & 0 & 0 \\ 0 & 1 & 0 \\ 0 & 0 & (\tau_1/\tau_2)^2 \end{pmatrix}, \kappa'_s = \infty, \kappa'_c = (\tau_1/\tau_2)^2 \tag{5}$$

where $\rho'_c$, $\kappa'_c$, $\rho'_s$ and $\kappa'_s$ are the necessitating materials of compressive and stretching regions, respectively. The obtained materials of the stretching region are named as nihility media since they have the competency of making total transmission condition . In other words, these media could be considered as a void since no reflection/ diffraction would be generated for the incident wave inside these media

## Numerical Results

To validate the concept, several innovative examples in 2D space were carried out by using COMSOL MULTIPHYSICS finite element solver. For all the performed simulations, the incident wave was taken to be an acoustic plane wave, which is illuminating under the frequency of $f = 4kHz$ with the amplitude of $A = 1$. Since the solving area is a rectangle with the sides of $16\lambda \times 16\lambda$, where $\lambda$ is the operative wavelength, the provided simulations exhibit both near field and far field behavior of the structure. Moreover, for all the proposed geometry of $R_1 = \tau_1 R(\phi)$, $R_2 = \tau_2 R(\phi)$ and $R_3 = \tau_3 R(\phi)$ constant values of $\tau_1 = 0.5$, $\tau_2 = 0.99$ and $\tau_3 = 1$ are selected, while the coefficients of $R(\phi)$ will achieve different values for each new case. Therefore, according to Eq.(5) and under selecting such values, the demanding materials will be achieved as:

$$\rho'_c = \rho_0 \begin{pmatrix} 1 & 0 & 0 \\ 0 & 1 & 0 \\ 0 & 0 & 0.25 \end{pmatrix}, \quad \rho'_s = \begin{pmatrix} 0 & 0 & 0 \\ 0 & \infty & 0 \\ 0 & 0 & \infty \end{pmatrix}, \quad \kappa'_c = 0.25\kappa_0, \quad \kappa'_s = \infty \tag{6}$$

The first example is dedicated to cylindrical cross-section concentrators with $R(\phi) = 3\lambda$ and its results are illustrated in Fig.3.

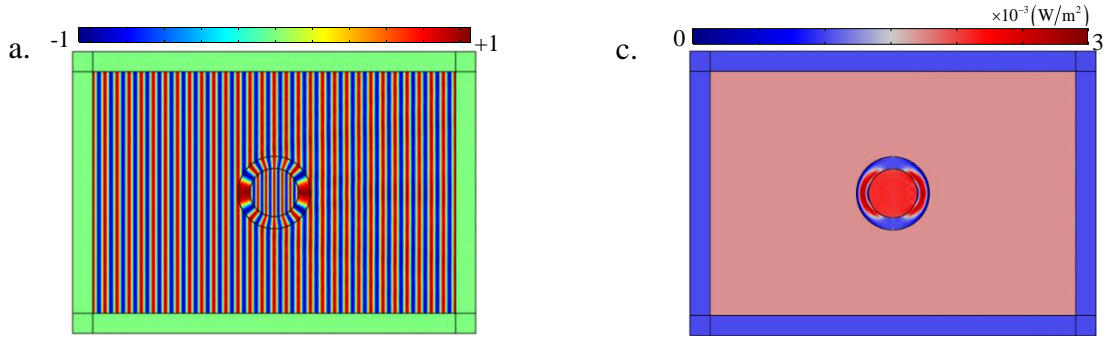

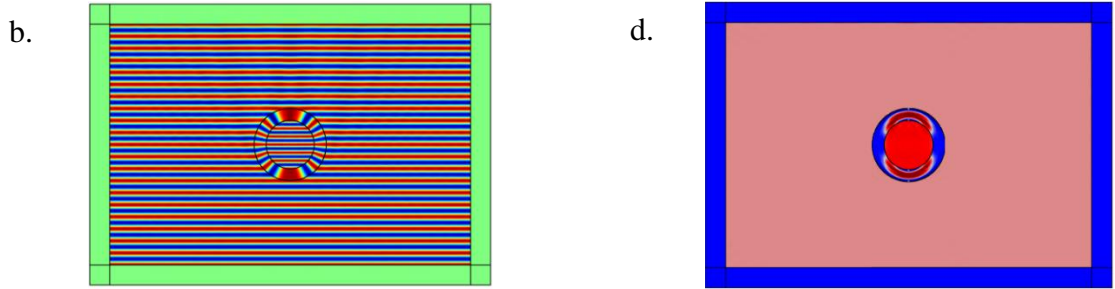

**Fig.3.** The pressure field of the cylindrical cross section of acoustic concentrator along the z- direction under three different incident angles of (a) $\theta = 0°$ (b) $\theta = 90°$ .(c-d) their corresponding acoustic intensity

As it can be seen from Fig.3(c) and Fig.3(d), the magnitude of the acoustic intensity, in the compressive region is perfectly enhanced regardless of the angle of incidence which are well abides with previously reported results in this area. However, to have full control of acoustic waves it might be necessary in some cases that the waves are being focused in an arbitrary cross-section. To this aim, several works had been performed previously, which their inhomogeneous and anisotropic materials limit their practical use. In addition to the complexity of the obtained materials, if the geometry of the concentrators is changed, one must recalculate the corresponding materials, which is time-consuming and also impractical for scenarios where reconfigurability is of utmost importance. Nevertheless, since the proposed materials of Eq.(6) are not restricted to any specific shapes, they could be used for any desired geometry. In other words, exploiting the materials of Eq.(6) would obviate the need for recalculating the necessitating materials after changing the shape of the device. To show this, assume that the contour coefficients of $R(\phi)$ in Eq.(4) are given in a way that a heart-shape concentrator is generated. By utilizing the obtained materials for each compressive and stretching regions, the results of the pressure field and acoustic intensity will be achieved as shown in Fig.4.

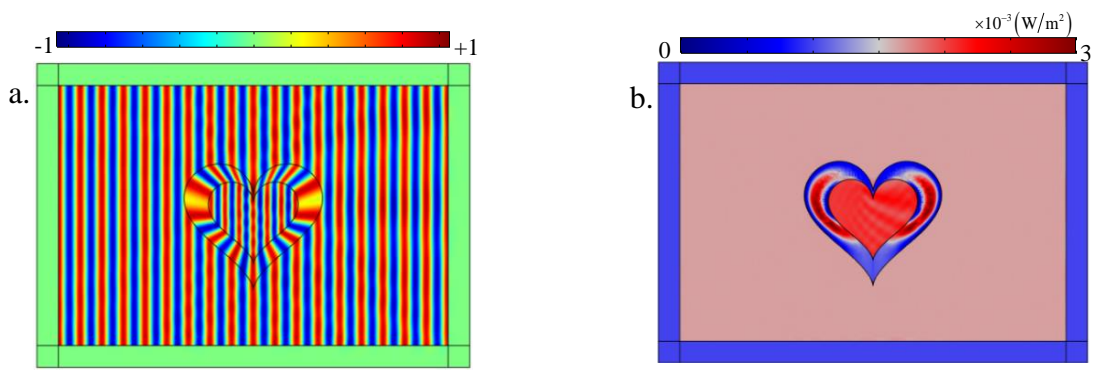

**Fig.4.** (a) The pressure field and (b) acoustic intensity of a heart-shape concentrator

As it is illustrated in Fig.4 (b), the pressure fields are well concentrated in the heart shape region with high acoustic intensity depicted in Fig.4 (b). In addition, it is evident that the incident wave does not scatter even for irregular shapes. To further demonstrate the independence of the proposed method to the geometry of the concentrators, the same material of Eq. (6) are used for other arbitrary cross-section concentrators and their results are depicted in Fig.5.

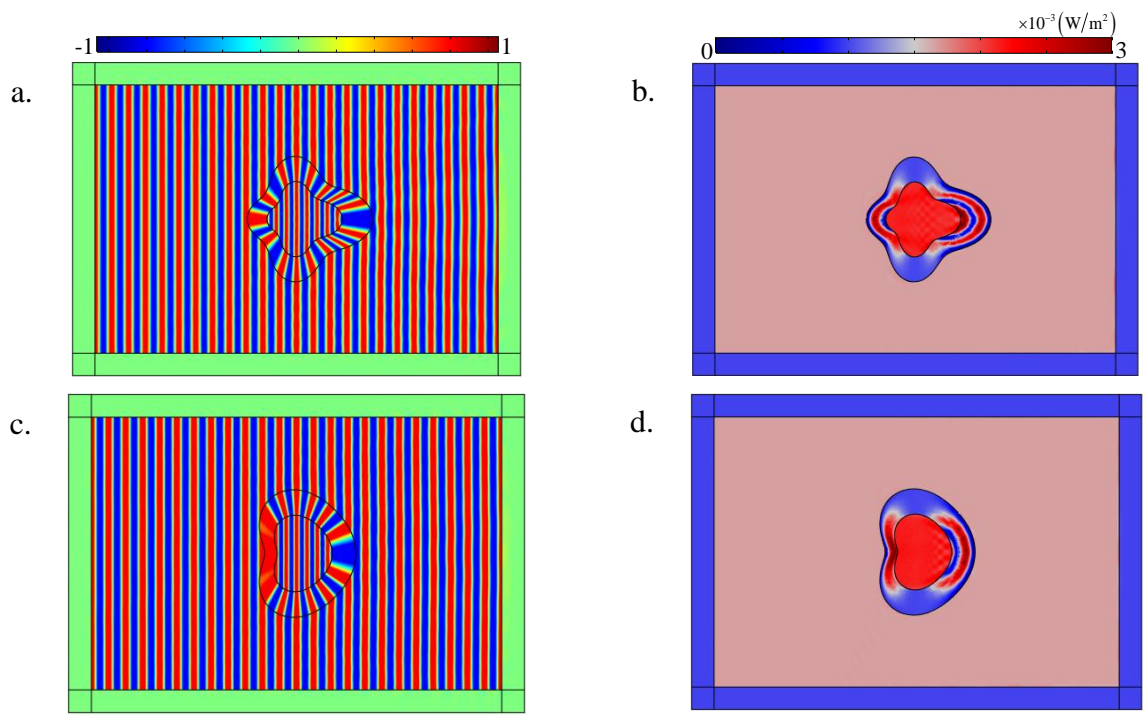

**Fig.5. (a),(c)** The pressure field of a arbitrary shape concentrators **(b),(d)** Their corresponding acoustic intensities.

As can be seen from Fig.5, different shape concentrators are achieved without changing the materials of Eq. (6) and their results exhibit strong agreement with their previous counterparts which were achieved by performing tedious calculations [12-15].

## Conclusion

In summary, for the first time, we have proposed a method in which any arbitrary shape acoustic concentrator will be achieved via naturally occurring materials. It was shown that the functionality of the proposed material which is an acoustic nihility media(ANM) is not restricted to the device geometry; hence, a constant ANM could be used for any shape concentrators. To validate the concept, several numerical simulations were performed , which their results were well abide by the theoretical predictions. We believe that the newly proposed method in this paper could pave the way towards the more unprecedented of acoustic manipulation especially in scenarios where the high concentration of sound waves is of utmost importance such as in sonar systems.